\documentclass[12pt]{article}

\usepackage{amsmath,amssymb,amsfonts,amsbsy}
\usepackage{cite}
\usepackage{graphicx}
\usepackage{wrapfig}

\usepackage{epsfig}

\begin{document}
\addcontentsline{toc}{subsection}{{Title of the article}\\
{\it B.B. Author-Speaker}}

\setcounter{section}{0}
\setcounter{subsection}{0}
\setcounter{equation}{0}
\setcounter{figure}{0}
\setcounter{footnote}{0}
\setcounter{table}{0}

\begin{center}
\textbf{COMPASS RESULTS ON GLUON POLARISATION FROM HIGH PT HADRON PAIRS}

\vspace{5mm}



L.~Silva\\
\begin{small}
On behalf of the COMPASS Collaboration.
\end{small}
\vspace{5mm}

\begin{small}
  \emph{LIP Lisboa} \\
  \emph{E-mail: lsilva@lip.pt}
\end{small}
\end{center}

\vspace{0.0mm} 

\begin{abstract}
One of the goals of the COMPASS experiment is the determination of the gluon
polarisation $\Delta G/G$, for a deep understanding of the spin structure
of the nucleon. In DIS the gluon polarisation can be measured via
the Photon-Gluon-Fusion (PGF) process, identified by open charm production or by selecting high $p_T$ hadron pairs in the final state. The data used for this work were collected by the COMPASS experiment
during the years 2002-2004, using a 160 GeV naturally polarised
positive muon beam scattering 
on a polarised nucleon target.
A new preliminary result of the gluon polarisation $\Delta G/G$ from high $p_T$ hadron pairs in events with $Q^2>1 \ (\mbox{GeV/}c)^2$ is
presented. In order to extract $\Delta G/G$, this analysis takes into
account the leading process $\gamma q$ contribution together with the PGF and
QCD Compton processes. A new weighted method based on a
neural network approach is used. A preliminary $\Delta G/G$ result for
events from quasi-real photoproduction ($Q^2<1 \ (\mbox{GeV/}c)^2$) is
also presented. 
\end{abstract}

\vspace{7.2mm}

\section{Introduction}
\label{intro}

The COMPASS experiment is located in the Super
Proton Synchrotron (SPS) accelerator at CERN. For a more complete description of the experimental apparatus the reader is
addressed to \cite{compass}. In 2007, the COMPASS
collaboration estimated the quark contribution to the nucleon spin with high precision \cite{comp.del_sigma},
using a NLO QCD fit with all world data available. This contribution confirms that approximately 1/3 of the nucleon spin is
carried by the quarks, as demonstrated by earlier experiments \cite{emc}.

The nucleon spin can be written as:
\begin{equation}
\frac{1}{2}=\frac{1}{2}\Delta \Sigma + \Delta G + L
\end{equation}

$\Delta \Sigma$ and $\Delta G$ are, respectively, the quark and gluon contributions to the nucleon spin and $L$ is the orbital angular momentum contribution coming from  from the quarks and gluons.

The aim of this analysis is to estimate the gluon polarisation, $\Delta G/G$, using the high
transverse momentum (high $p_T$) hadron pairs sample. The analysis is
performed in two complementary kinematic regions: $Q^2 <1 \ (\mbox{GeV}/c)^2$
(low $Q^2$) and $Q^2 >1 \ (\mbox{GeV}/c)^2$ (high $Q^2$) regions. The
present work is mainly focused on the analysis for high $Q^2$. However,
the analysis for the low $Q^2$ region is summarised in sec.~\ref{sec:lowq2}.

For completeness, the slides of the presentation can be found in  \cite{presentation}.

\section{Analysis Formalism}
\label{sec:analysis}

Spin-dependent effects can be measured experimentally using the helicity asymmetry 

\begin{equation}
  A_{\rm LL} = \frac{\Delta \sigma}{2 \sigma}=
  \frac{\sigma^{\uparrow \Downarrow} - \sigma^{\uparrow \Uparrow}
  }{\sigma^{\uparrow \Downarrow} + \sigma^{\uparrow \Uparrow} } \label{eq:asy}
\end{equation}
defined as the ratio of polarised ($\Delta \sigma$) and unpolarised ($\sigma$)
cross sections. $\uparrow \Uparrow$ and $\uparrow \Downarrow$ refer to
the parallel and anti-parallel spin helicity configuration of the beam lepton ($\uparrow$) with respect to the target nucleon ($\Uparrow$ or $\Downarrow$).

According to the factorisation theorem, the (polarised) cross
sections can be written as the convolution of the (polarised) parton distribution
functions, ($\Delta$)$q_i$, the hard scattering partonic cross section,
($\Delta$)$\hat{\sigma}$, and the fragmentation function $D_f$.

\begin{wrapfigure}[11]{r}{10cm}
\vspace{-30pt}
\begin{center}
\mbox{\epsfig{figure=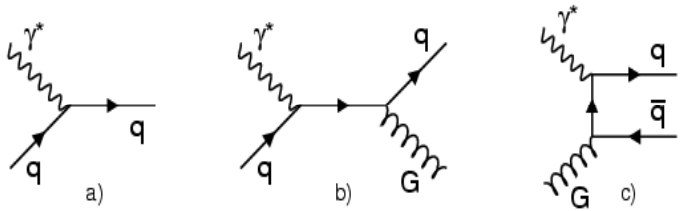,width=10cm,height=3.5cm}
}
\end{center}
\caption{The contributing processes: a) DIS LO, b) QCD Compton and c) Photon-Gluon Fusion.}
\label{fig:procs}
\end{wrapfigure}

The gluon polarisation is measured directly via the Pho\-ton-Gluon
Fusion process (PGF); which allows to probe the spin of the gluon
inside the nucleon. To tag this process directly in DIS a high $p_T$ hadron pairs data sample is used to calculate the
helicity asymmetry. Two other processes
compete with the PGF process in leading order QCD approximation, namely
the virtual photo-absorption leading order (LO) process and the gluon
radiation (QCD Compton) process. In Fig.~\ref{fig:procs} all contributing processes are depicted.

The helicity asymmetry for the high $p_T$ hadron pairs data sample can
thus be schematically written as:
\begin{equation}
A_{\rm LL}^{2h}(x_{Bj})= R_{\rm PGF} \, a_{\rm LL}^{\rm PGF}\frac{\Delta G}{G}(x_G) +
R_{\rm LO} \, D \, A_1^{\rm LO}(x_{Bj}) + R_{\rm QCDC} \, a_{\rm LL}^{\rm QCDC}
A_1^{\rm LO}(x_C) \label{eq:allmain}
\end{equation}

The $R_i$ (the index $i$ refers to the different processes) are the fractions of each process. $a_{\rm LL}^i$ represents the
partonic cross section asymmetries, $\Delta\hat{\sigma}^i/\hat{\sigma}^i$, (also known as analysing power). $D$
is the depolarisation factor~\footnote{The depolarisation factor is the fraction of the muon beam polarisation transferred to the virtual photon.}. The virtual photon asymmetry $A_1^{\rm LO}$ is defined as $A_1^{\rm LO} \equiv \frac{\sum_i e_i^2 \Delta q_i}{\sum_i e_i^2 q_i}$.

To extract $\Delta G/G$ from eq.~(\ref{eq:allmain}) the contribution
from the physical background processes LO and QCD Compton needs to be
estimated. This is done using Monte Carlo (MC) simulation to calculate $R_i$
fractions and $a_{\rm LL}^i$. The virtual photon asymmetry $A_1^{\rm LO}$ is
estimated using a parametrisation based on the $A_1$ asymmetry
 of the inclusive data \cite{compassrho}. Therefore a similar equation to
(\ref{eq:allmain}) can be written to express the inclusive asymmetry
 of a data sample, $A_{\rm LL}^{incl}$.

Using eq.~(\ref{eq:allmain}) for the high $p_T$ hadron pairs sample
and a similar eq. for the inclusive sample the following expression is obtained:

\begin{eqnarray}
\frac{\Delta G}{G}(x_G^{av}) &=& \frac{A_{\rm LL}^{2h}(x_{Bj})+A^{corr} }{\beta}\nonumber\\
A^{corr}&=& - A_1(x_{Bj})D \frac{R_{\rm LO}}{R_{\rm LO}^{incl}} - A_1(x_C)
\beta_1 + A_1(x_C')\beta_2
 \label{eq:form:gluon}
\end{eqnarray}
and
\begin{eqnarray}
\beta_1 &=& \frac{1}{R_{\rm LO}^{incl}}\bigg[a_{\rm LL}^{\rm QCDC} R_{\rm QCDC} -a_{\rm LL}^{incl,{\rm QCDC}} R_{\rm QCDC}^{incl}\frac{R_{\rm LO}}{R_{\rm LO}^{incl}}\bigg]\nonumber\\
\beta_2 &=& a_{\rm LL}^{incl,{\rm QCDC}}\frac{R_{\rm QCDC}^{incl}}{R_{\rm LO}^{incl}}\frac{R_{\rm QCDC}}{R_{\rm LO}^{incl}}\frac{a_{\rm LL}^{\rm QCDC}}{D}\nonumber\\
\label{eq:form:betas}
\alpha_1 &=& a_{\rm LL}^{\rm PGF} R_{\rm PGF} - a_{\rm LL}^{incl,{\rm PGF}} R_{\rm LO}\frac{R_{\rm PGF}^{incl}}{R_{\rm LO}^{incl}} \\
\label{eq:form:alfa:betas}
\alpha_2 &=& a_{\rm LL}^{incl,{\rm PGF}}R_{\rm QCDC}\frac{R_{\rm PGF}^{incl}}{R_{\rm LO}^{incl}}\frac{a_{\rm LL}^{\rm QCDC}}{D}\nonumber\\
\beta &=& \alpha_1-\alpha_2.\nonumber
\end{eqnarray}

The term $A^{corr}$ comprises the correction due to the other two
processes, namely the LO and the QCD Compton processes. $\alpha_1$, $\alpha_2$,
$\beta_1$, $\beta_2$, $x_C$, $x_C'$ and $x_G^{av}$ are estimated using high $p_T$ and
inclusive MC samples.

\section{Data Selection}
\label{sec:data}

Data from 2002 to 2004 years is used. The selected events have an
interation vertex containing an incoming muon beam and a scattered muon. As mentioned in sec.~\ref{sec:analysis} the data
samples are divided into two data sets: the high $p_T$ hadron pairs and
the inclusive data samples.

Both data sets have the $Q^2>1 \ (\mbox{GeV}/c)^2$ kinematic cut
  applied. Another cut is applied on the fraction of energy taken by
  the virtual photon, $y$: $0.1 < y < 0.9$. These cuts described previously are used to select the
  {\em inclusive sample}.

In the high $p_T$ hadrons data sample, besides the inclusive selection,
events with (at least) two outgoing high $p_{T}$ hadrons are considered. These so-called hadron candidates must
fulfill the following requirement: the two hadrons with the highest
transverse momentum must have $p_T > 0.7 \ \mbox{GeV/c}$. This
requirement constitutes the  high $p_T$ cut.  All these cuts
additionally correspond to the {\em high $p_T$ sample} selection.

\section{Monte Carlo simulation}
\label{MC}

Important information to be used in
the $\Delta G/G$ extraction is obtained from MC
simulation. In this analysis it is fundamental that the simulation
describes well the experimental data. Two MC samples were
produced: one for the high $p_T$ sample and another for the inclusive 
sample, to  estimate the terms in the set of eq.~(\ref{eq:form:betas}).

\begin{wrapfigure}[35]{r}{12cm}
\vspace{-7pt}
\centerline{
\includegraphics[clip,width=0.218\textwidth]{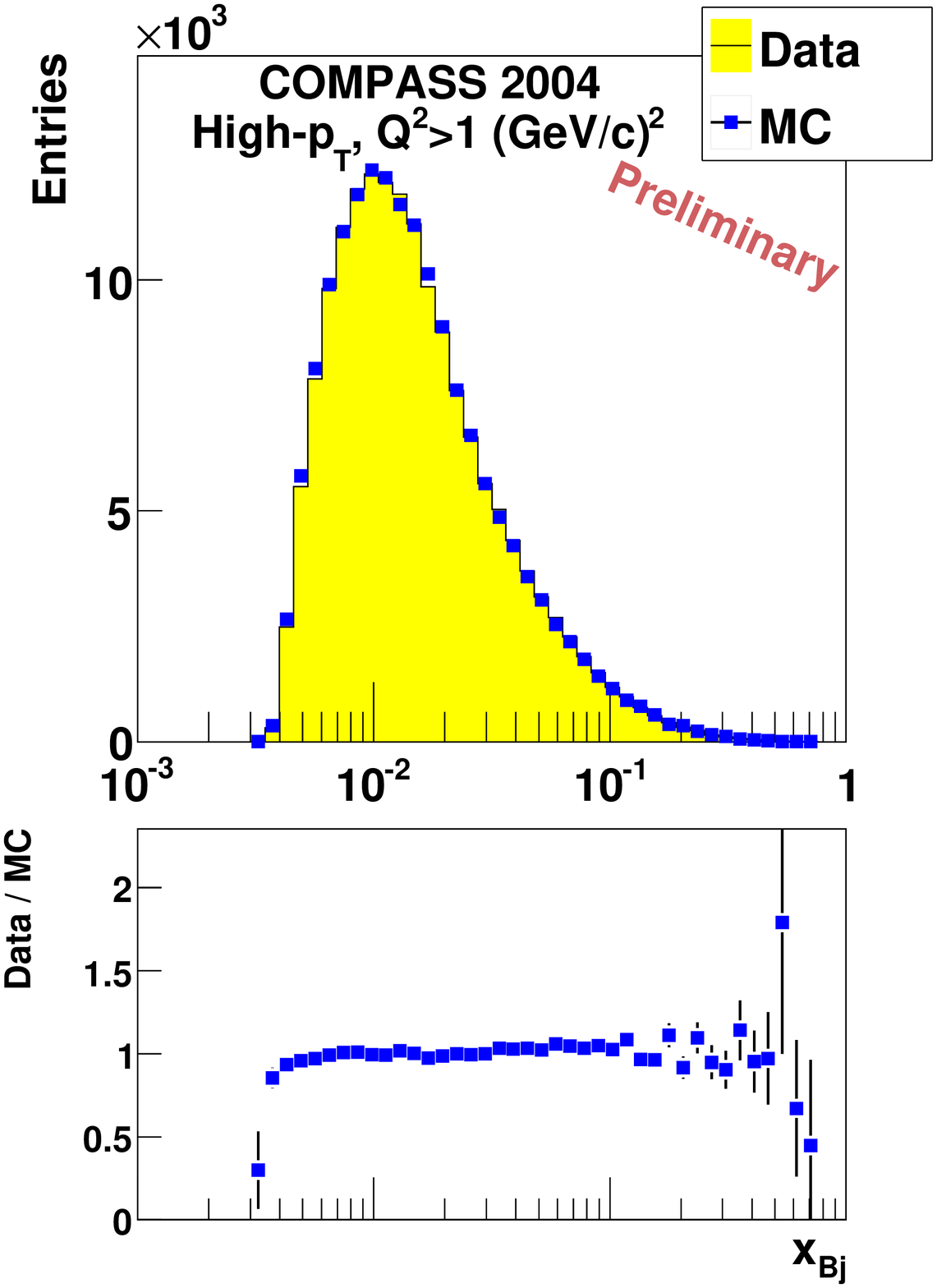}
\includegraphics[clip,width=0.218\textwidth]{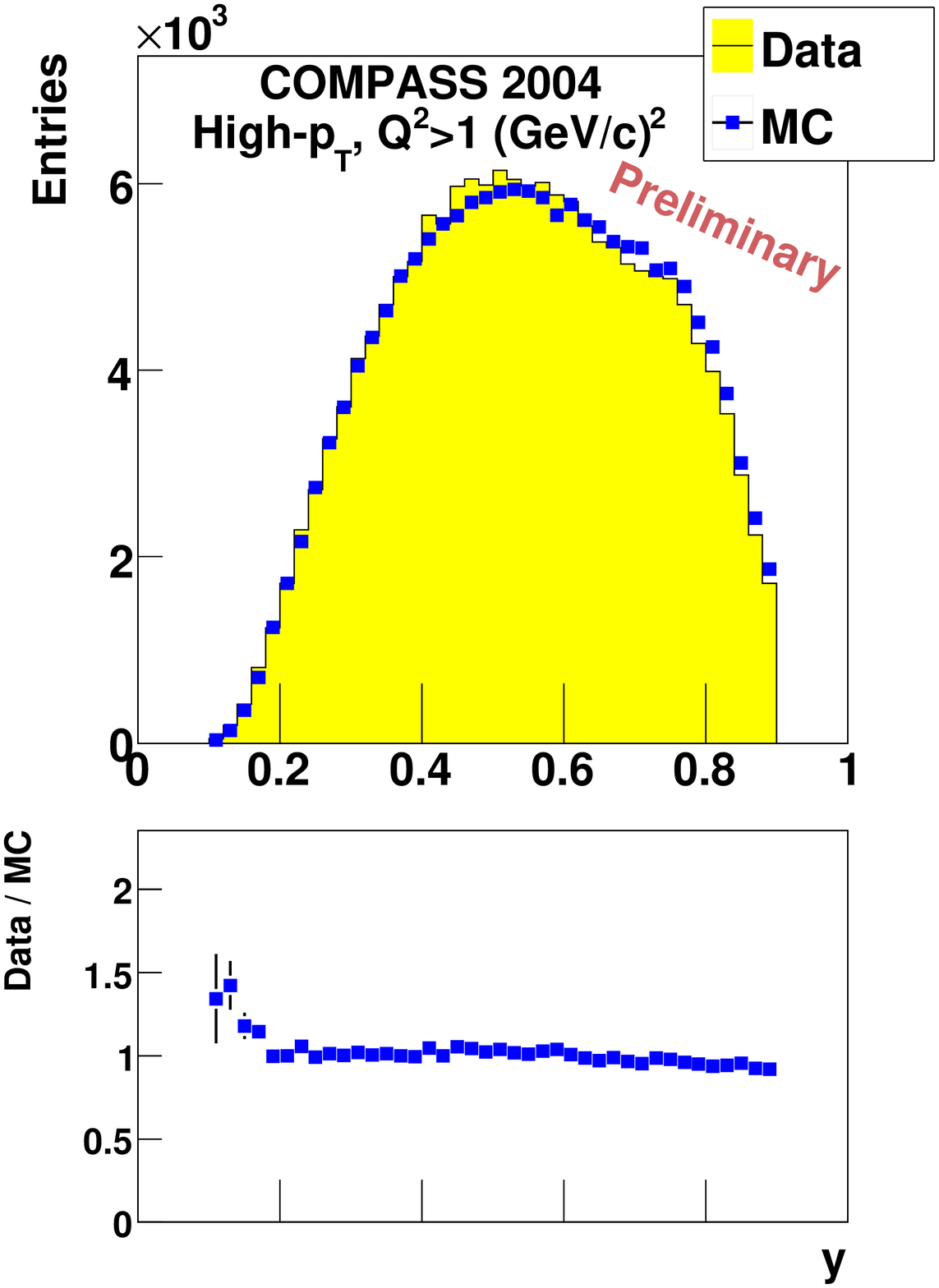}
\includegraphics[clip,width=0.218\textwidth]{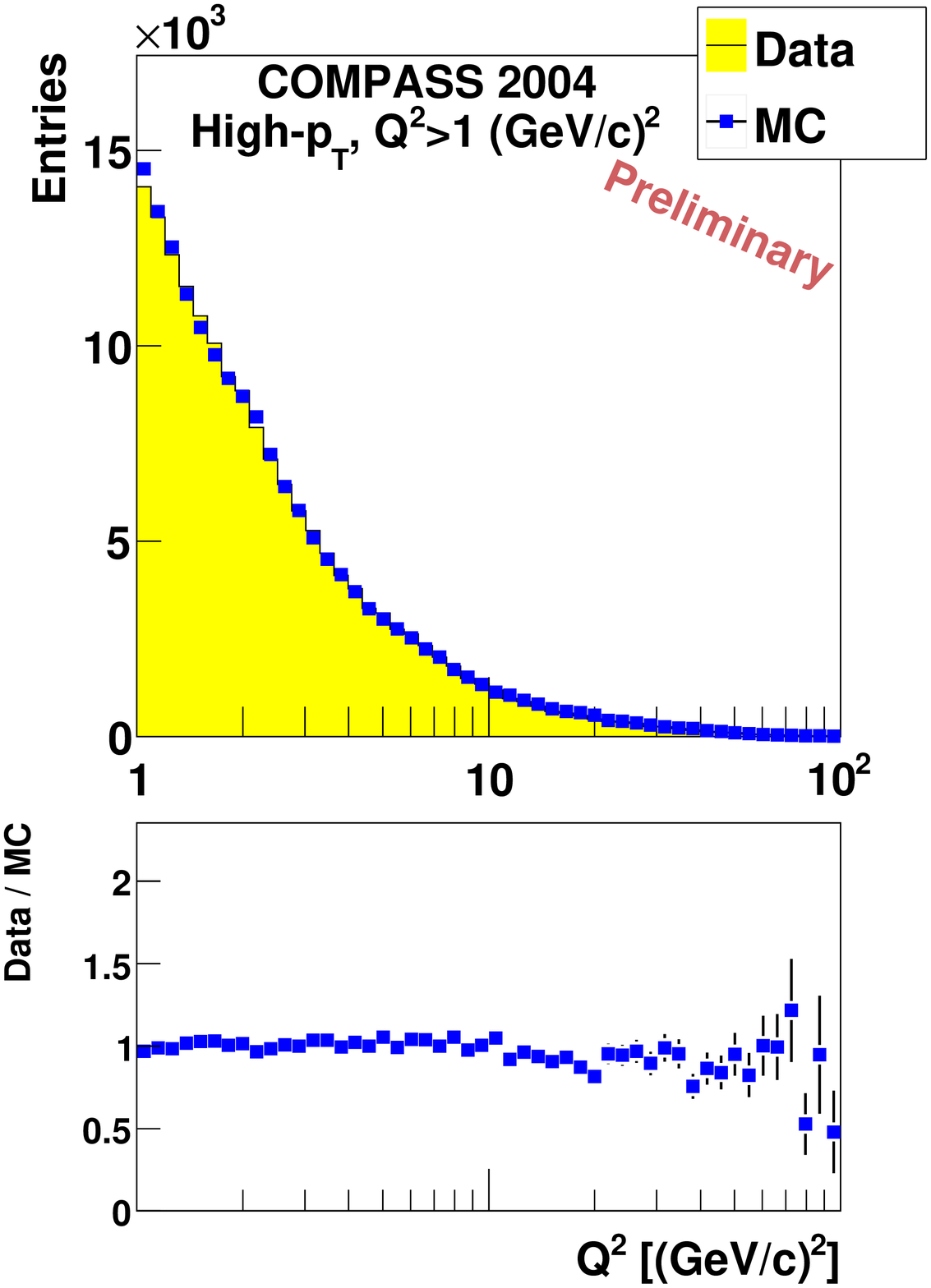}}
\centerline{
\includegraphics[clip,width=0.218\textwidth]{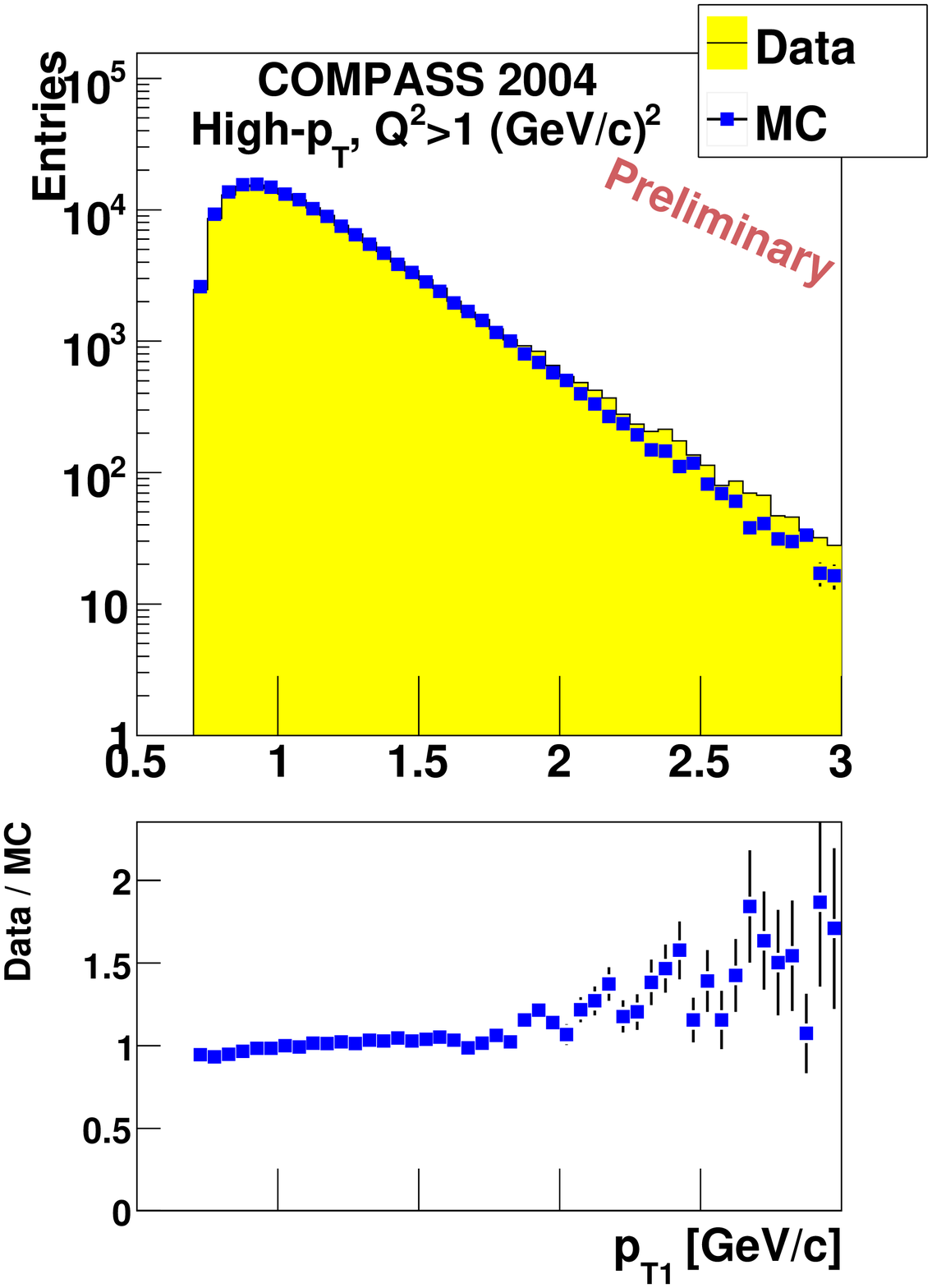}
\includegraphics[clip,width=0.218\textwidth]{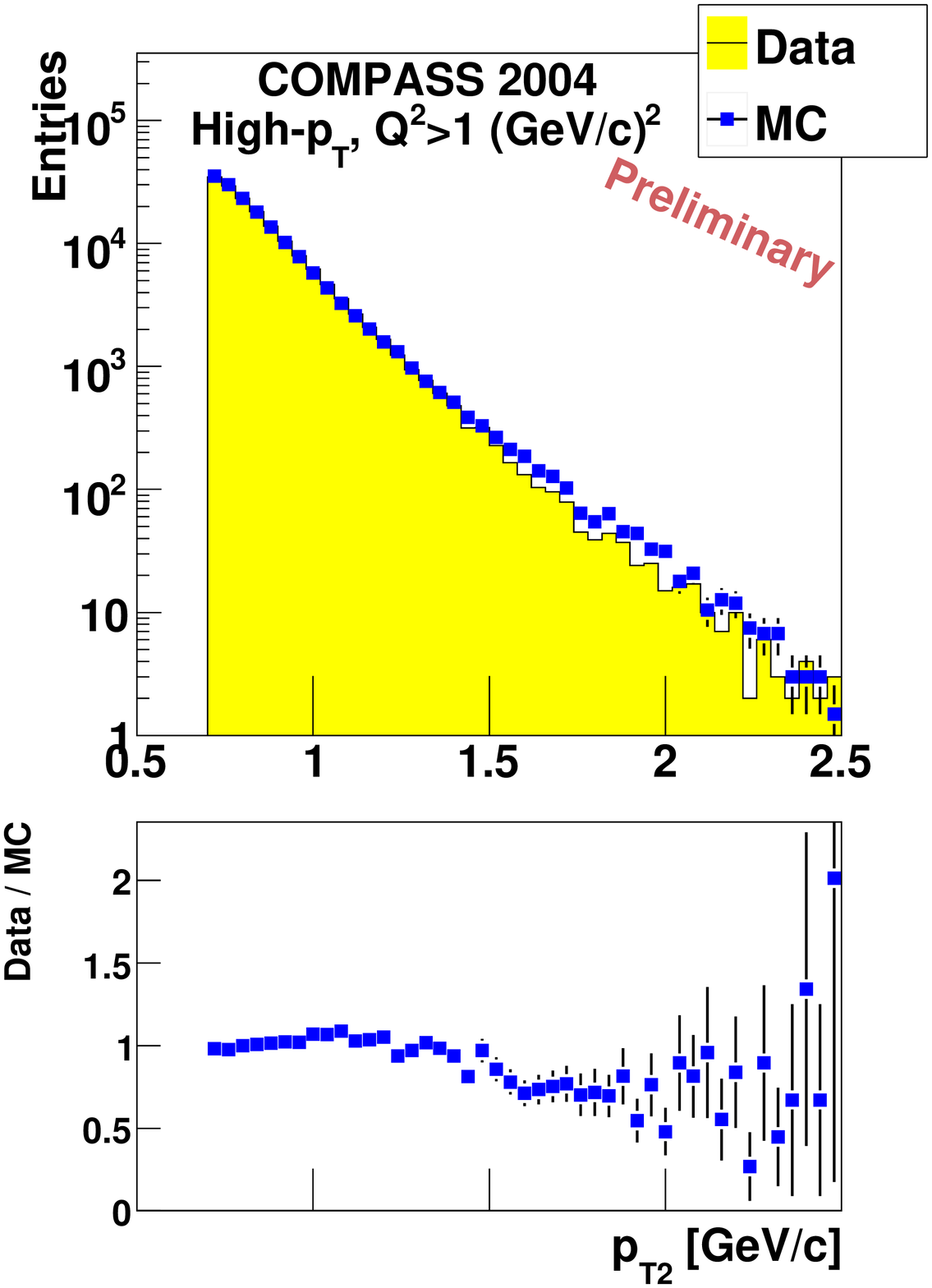}
\includegraphics[clip,width=0.218\textwidth]{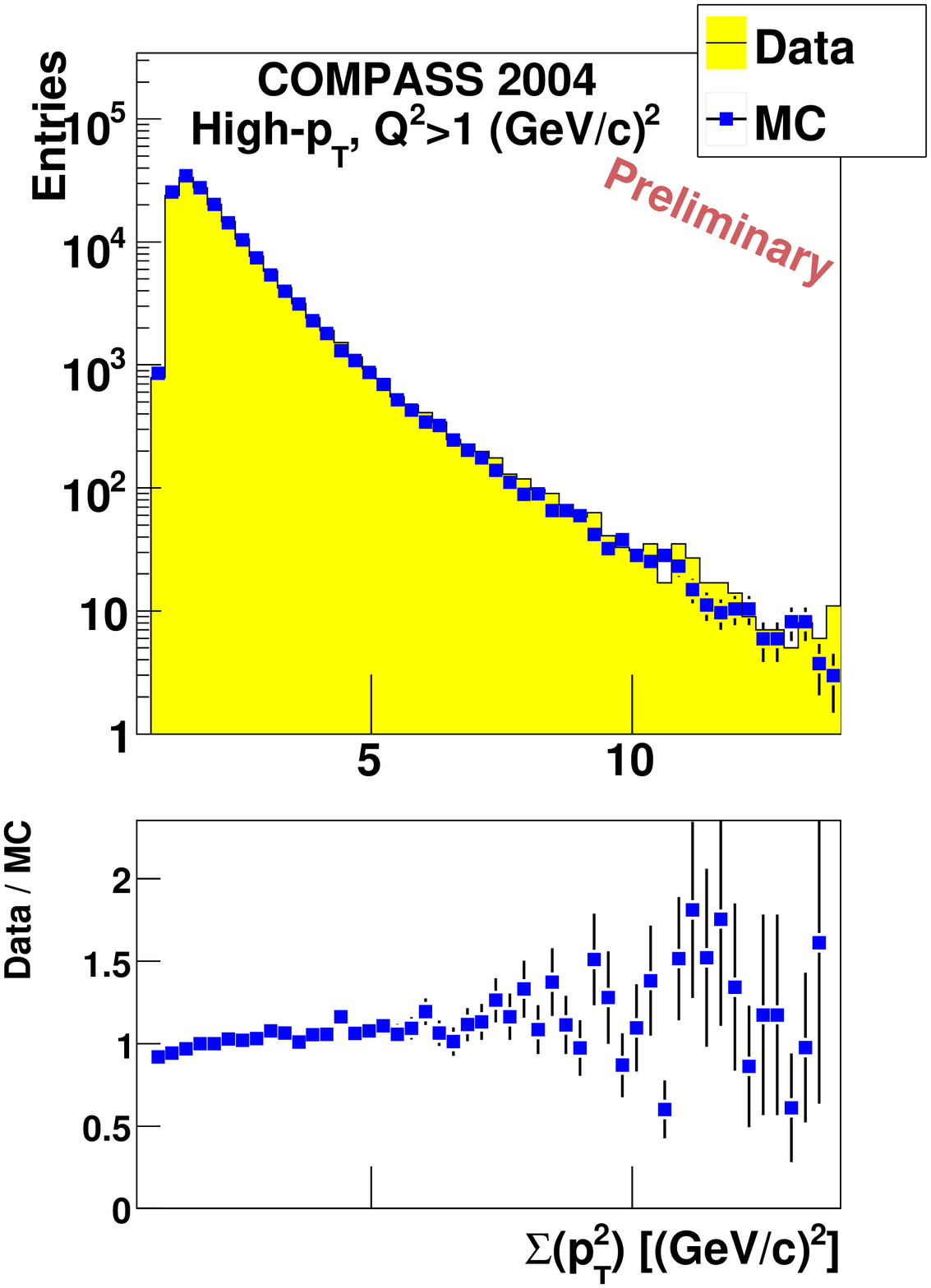}}
\centerline{
\includegraphics[clip,width=0.218\textwidth]{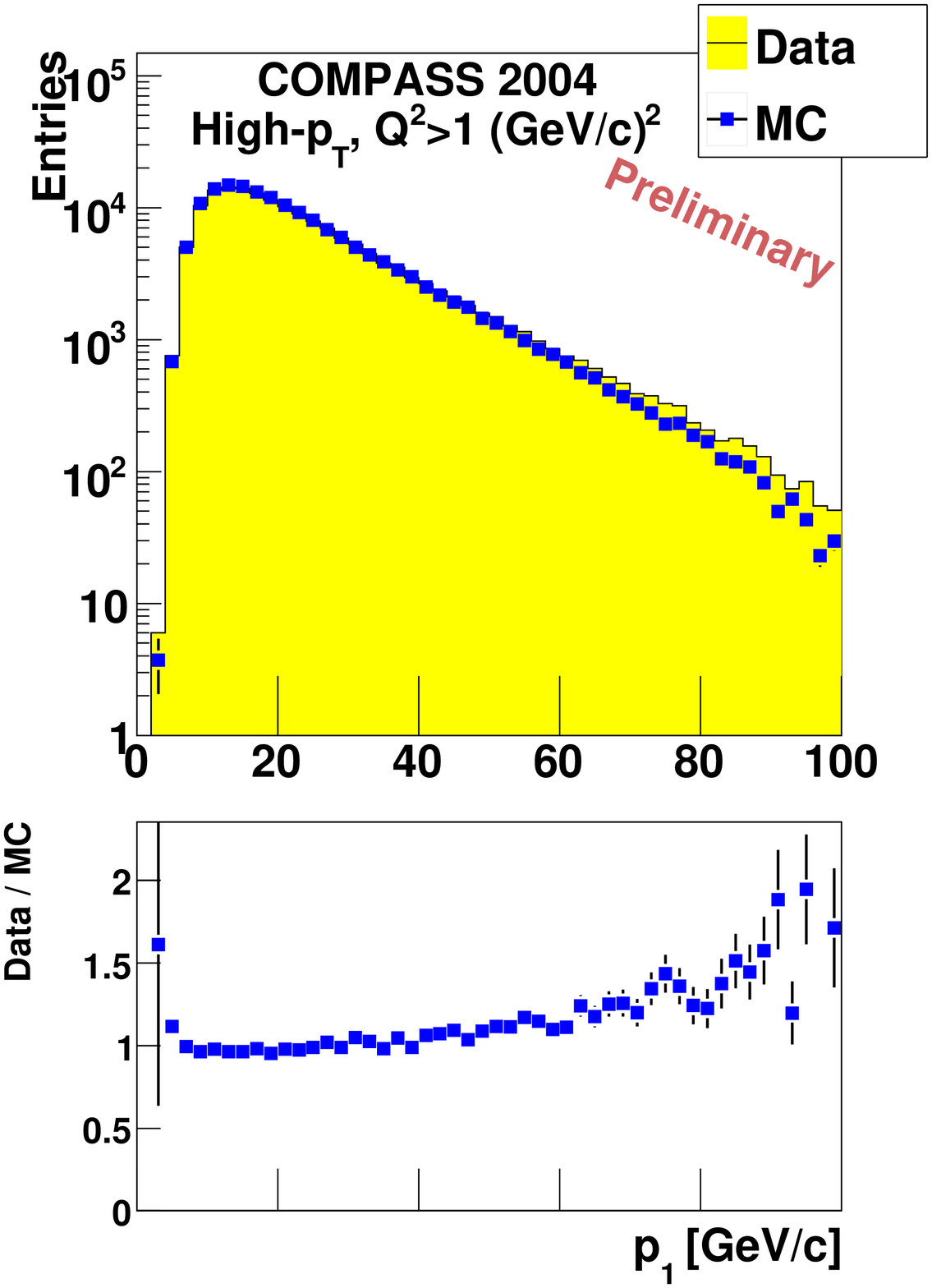}
\includegraphics[clip,width=0.218\textwidth]{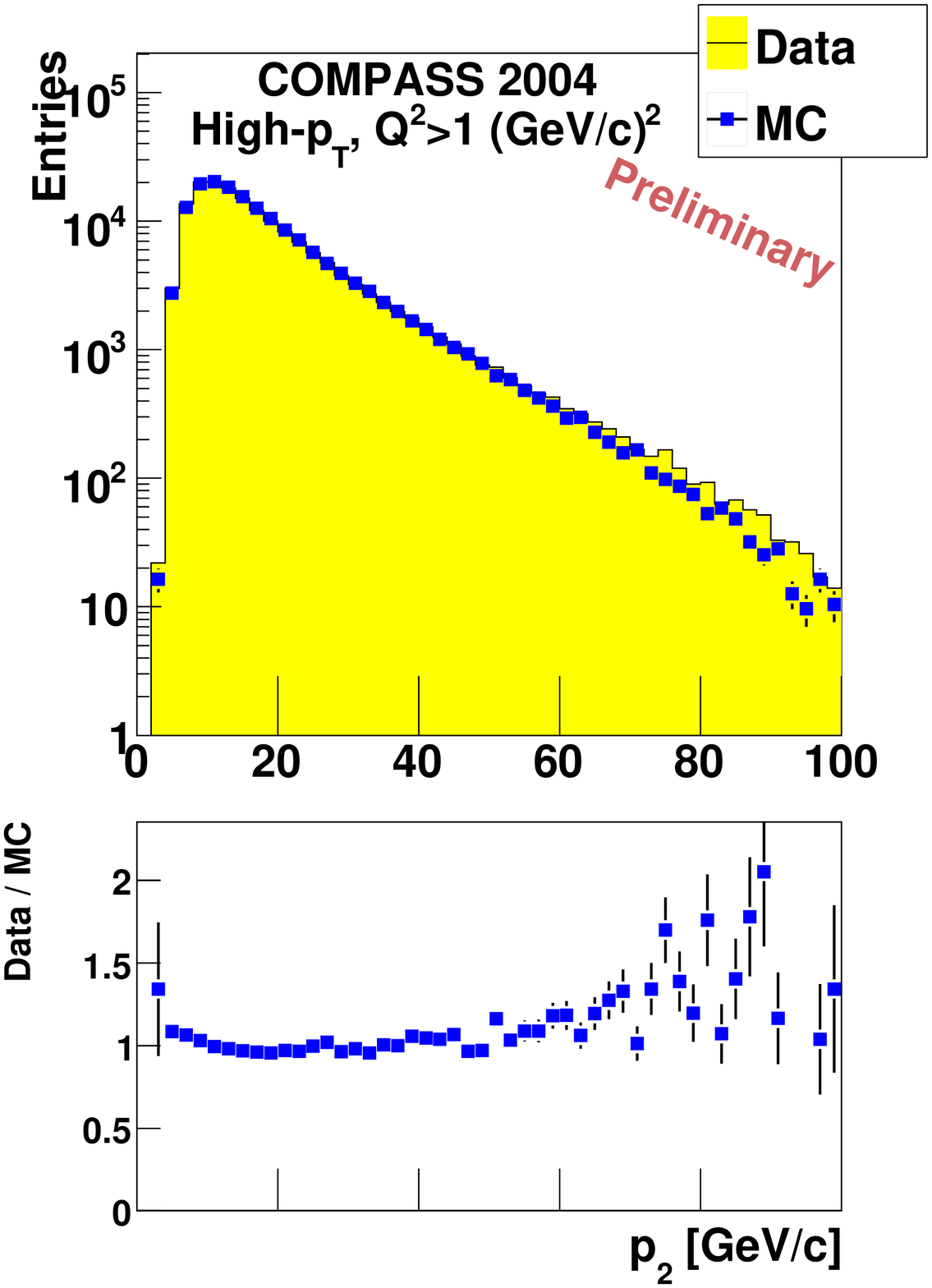}
\includegraphics[clip,width=0.218\textwidth]{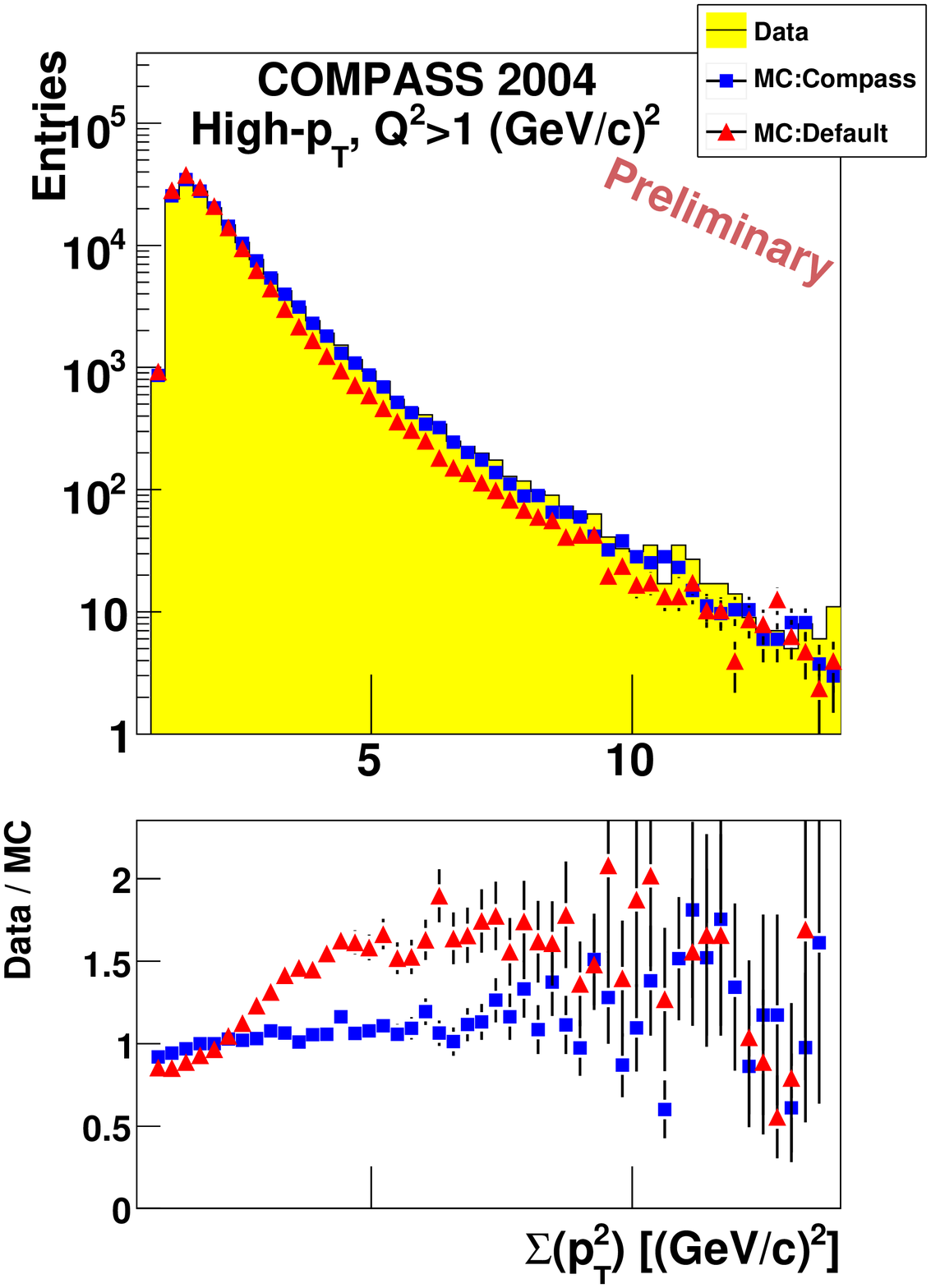}}
\caption{Comparison between data and MC simulations -- The
  distributions and ratios of Data/MC for: inclusive variables:
  $x_{Bj},\, Q^{2},\, y$ (1st row). For hadrons $p_T$ (2nd raw). For
  hadron momenta (3rd row), and also the comparison of MC with LEPTO
  default tuning.}
\label{fig:mc}
\end{wrapfigure}

The MC production comprises
three steps: first the events are generated, then the particles pass
through a simulated spectrometer using a program based on GEANT3 \cite{geant} and finally the events
are reconstructed using the same procedure applied to real data.

For the first step the LEPTO 6.5 \cite{Ingelman:1996mq} 
event generator is used together with a leading order para\-metrisation of the
unpolarised parton distributions. The MRST04LO set of parton
distributions is used in a fixed-flavour scheme generation. This set
of parton distributions has a good description of $F_2$ in the COMPASS kinematic region.

NLO
corrections are simulated partially by including gluon radiation in the initial and
final states (parton shower -- PS).

The fragmentation is based on the Lund
string model \cite{Andersson:89} implemented in JETSET \cite{Sjostrand:1985ys}. In this model the
probability that a fraction $z$ of the available energy will be
carried by a newly created hadron is expressed by the Lund symmetric function
$f(z)=z^{-1}(1-z)^a e^{-b m^2_\perp /z}$, with $m_\perp^2 = m^2 +
p_\perp^2$, where $m$ is the hadron mass.

To improve the agreement between MC
and data, the parameters ($a$,$b$) in the fragmentation function are modified from their default values (0.3,0.58) to (0.6, 0.1).

The transverse momentum of the hadrons, $p_T$, at the fragmentation level is
given by the sum of the $p_T$ of each hadron quarks. Then the
$p_T$ of the newly created hadrons is described by three steering
paramrters JETSET parameters: PARJ(21), PARJ(23) and PARJ(24). The default values of these
three parameters are (0.36, 0.01, 2.0), and were modified to (0.30, 0.02, 3.5). 

The remarkable agreement of the MC simulation with the data is illustrated
in Fig. \ref{fig:mc}; this figure shows the data--MC comparison of the kinematic variables: $x_{Bj}$, $y$ and $Q^2$ (1st row), the hadronic variables, $p_T$ for the
leading and sub-leading hadrons, together with the sum of
$p_T^2$, i.e. $\sum p_{T1}^2 +p_{T2}^2$ (2nd row), and the
momentum $p$ of those hadrons (3rd row), also two comparisons of the $\sum p_T^2$ variable one using the COMPASS
tuning and another using the default LEPTO tuning. In this example, it
is evident that the COMPASS tuning describes better our data sample than the LEPTO default one.

\section{The $\Delta G/G$ extraction method}
\label{deltag}
In the original idea of the high $p_T$ analysis, the selection was
based on a very tight set of cuts to suppress LO and QCD
Compton. This situation results in a dramatic loss of statistics. A
new approach was found, in which a loose set of cuts 
applied, combined with the use of a neural network \cite{Sulej:2007zz} to
assign a probability to each event to be originated from each of
the three processes. The main goal of this method is to enhance the PGF
process in the events sample, which accounts for the gluon contribution to the nucleon
spin. 

The neural network is trained using MC samples. In this way the neural
network is able to learn about the three processes in order to be
disentangled. A parametrisation of the variables $R_i$, $x^i$ and $a^i_{LL}$ for
each process type are estimated by the neural network using as input 
the kinematic variables: $x_{Bj}$ and $Q^{2}$, and the hadronic
variables: $p_{T1}$, $p_{T2}$, $p_{L1}$, and $p_{L2}$.

\begin{wrapfigure}[14]{r}{10cm}
\vspace{-10pt}
\centerline{
  \includegraphics[clip,width=0.3\textwidth]{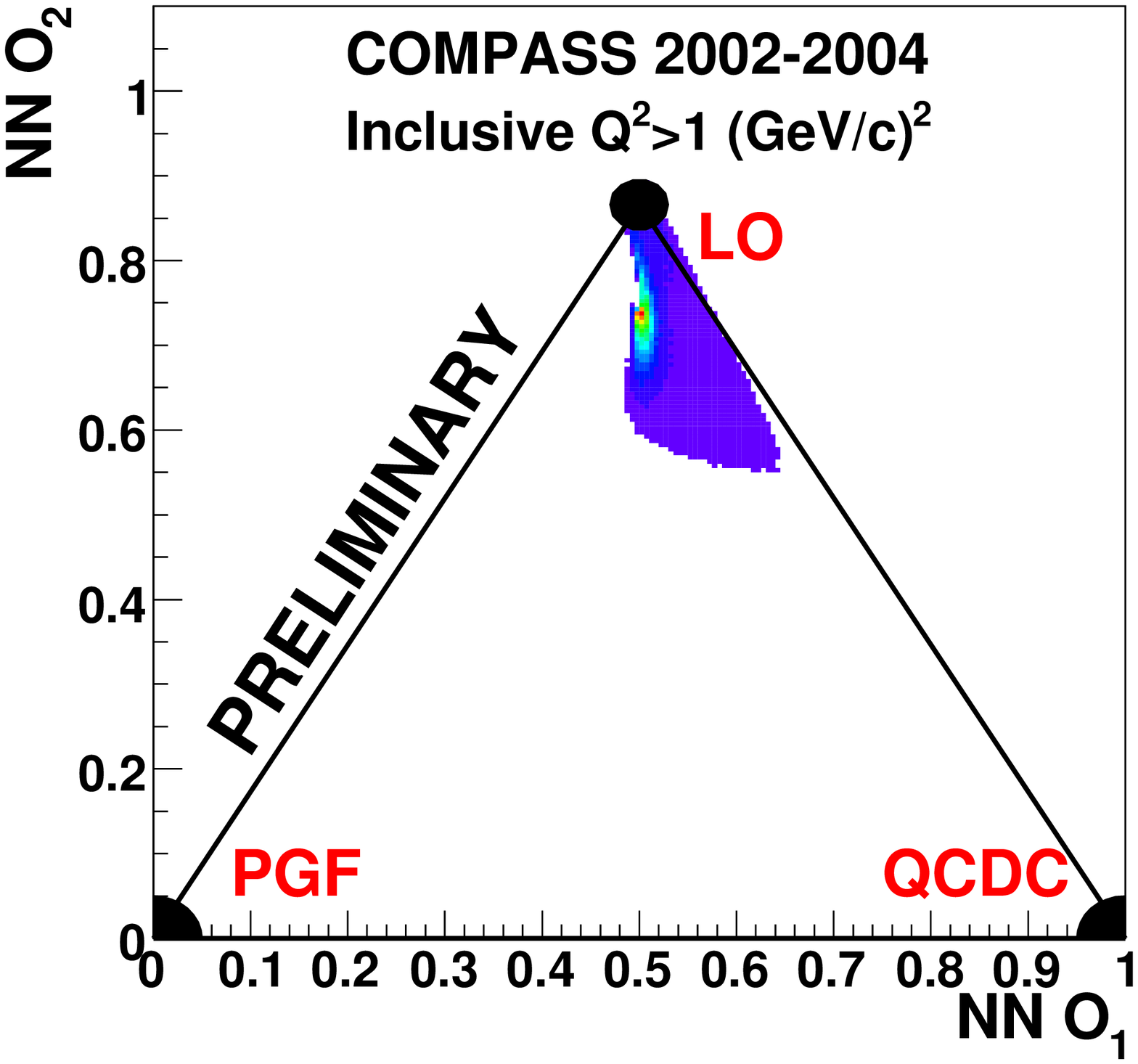}
  \includegraphics[clip,width=0.3\textwidth]{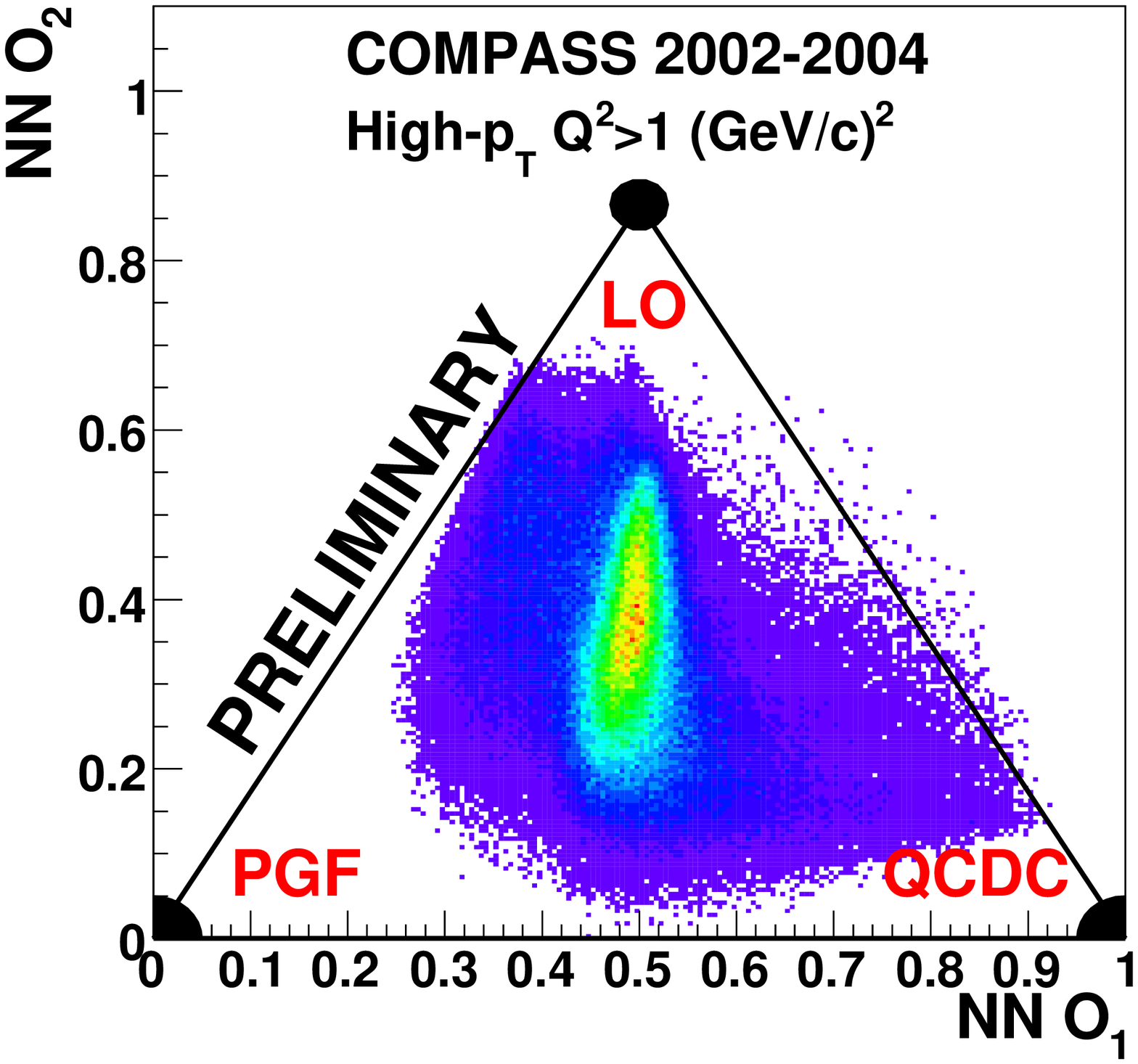}}
\caption{2-d output of neural network for estimation that the given  event is
PGF, QCDC or LO; (left) for the inclusive
sample and (right) for high $p_{T}$ sample.} \label{fig:form:2dsel}
\end{wrapfigure}

As the fractions of the three processes sum up to unity, we
need two variables to parameterise them: $o_1$ and $o_2$. The relations between the two neural network outputs $o_{1}$ and $o_{2}$ and
the fraction are $R_{\rm PGF} = 1-o_{1}-1/\sqrt{3}\cdot o_{2}$,
$R_{\rm QCDC}= o_{1}- 1/\sqrt{3}\cdot o_{2}$  and $R_{\rm LO}=  2/ \sqrt{3} \cdot o_{2}$.

A statistical weight is constructed for each event based on these 
probabilities. In this way we do not need to remove events that most likely do
not came from PGF processe, because the weight will naturally reduce their
contribution in the gluon polarisation, thus enhancing the sample of
events thathave a PGF likelihood.

The resulting neural network outputs for the fractions are presented in Fig.~\ref{fig:form:2dsel} in a 2-dimensional plot. The triangle
limits the region where all fractions are positive. For the
inclusive sample the average value of $o_2$ is quite large, which
means that the LO process is the dominant one. The situation is
different for the high $p_{T}$ sample, in which the average outputs are
$\langle o_1 \rangle \approx$ 0.5 and $\langle o_2 \rangle
\approx$ 0.35. Note also that the spread along $o_{2}$ is larger
than along $o_1$.

This means that the neural network is able to select a region where the contribution of PGF and QCDC is significant compared to LO, although it
can not easily distinguish between the PGF and QCDC processes
themselves.

\section{High $p_T$ hadron pair analysis for low $Q^2$ region }
\label{sec:lowq2}

\begin{wrapfigure}[23]{r}{8cm}
\vspace{-15pt}
\centerline{
\includegraphics[width=0.25\textwidth]{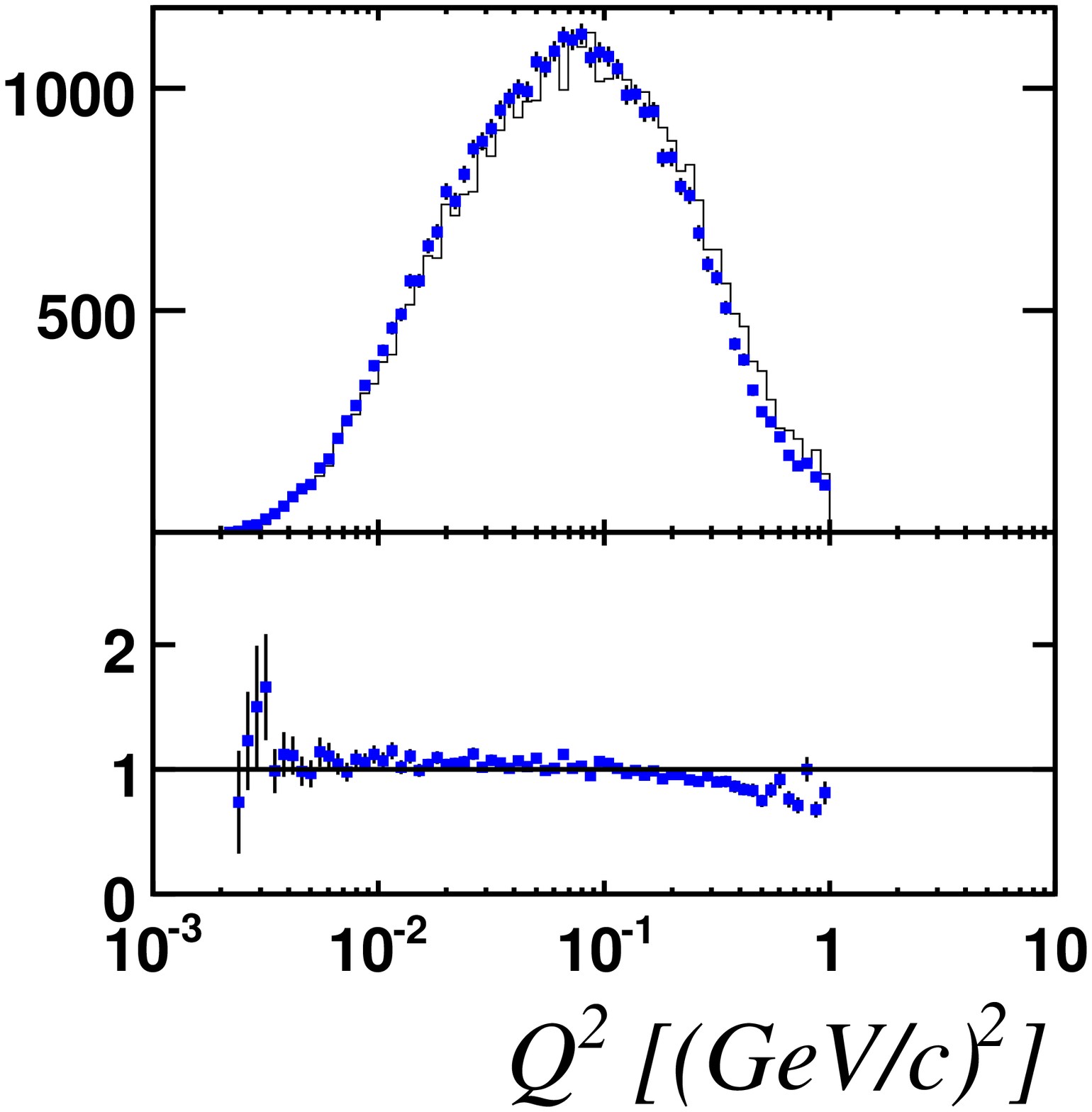}
\includegraphics[width=0.25\textwidth]{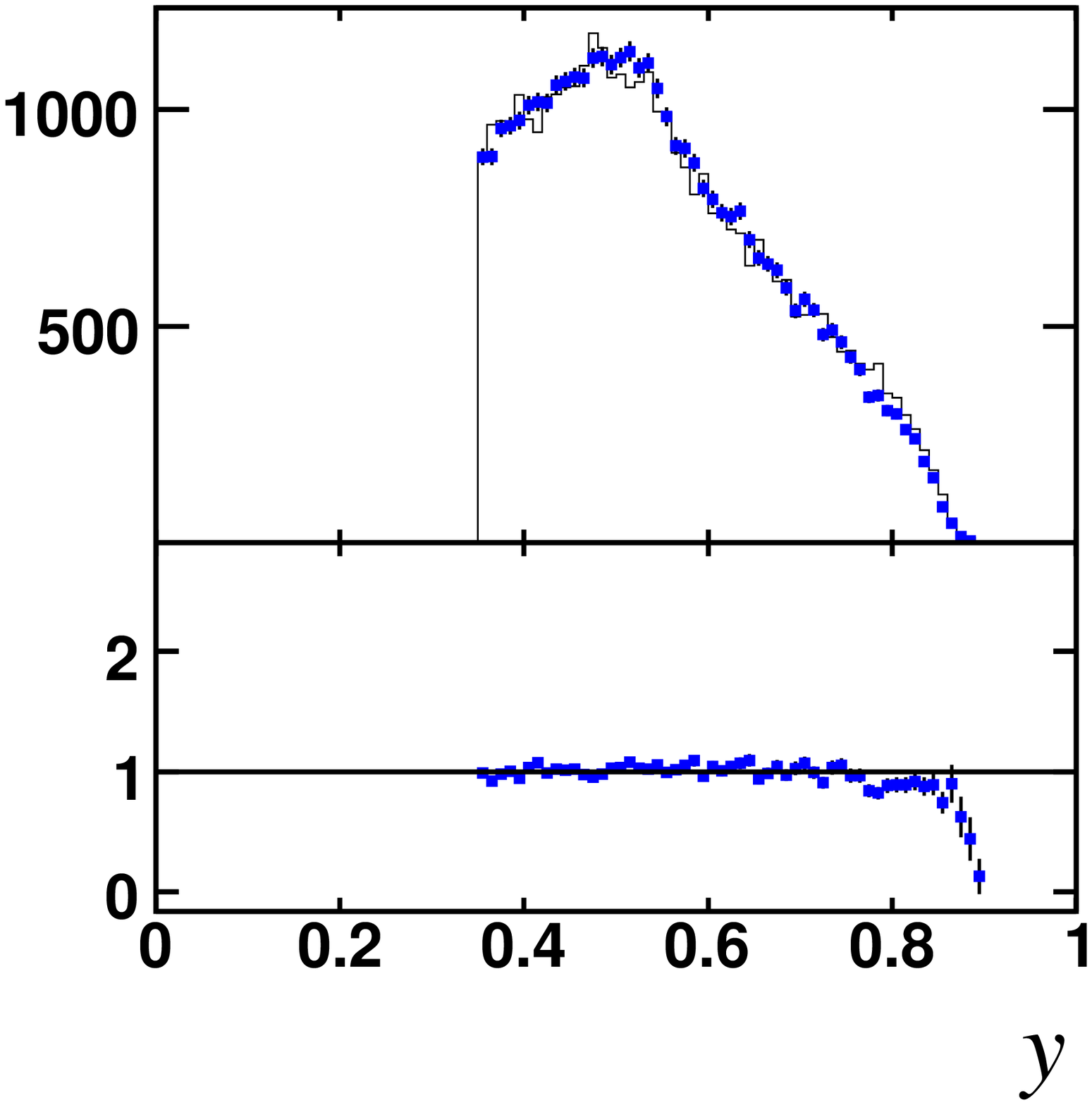}}
\centerline{
\includegraphics[width=0.25\textwidth]{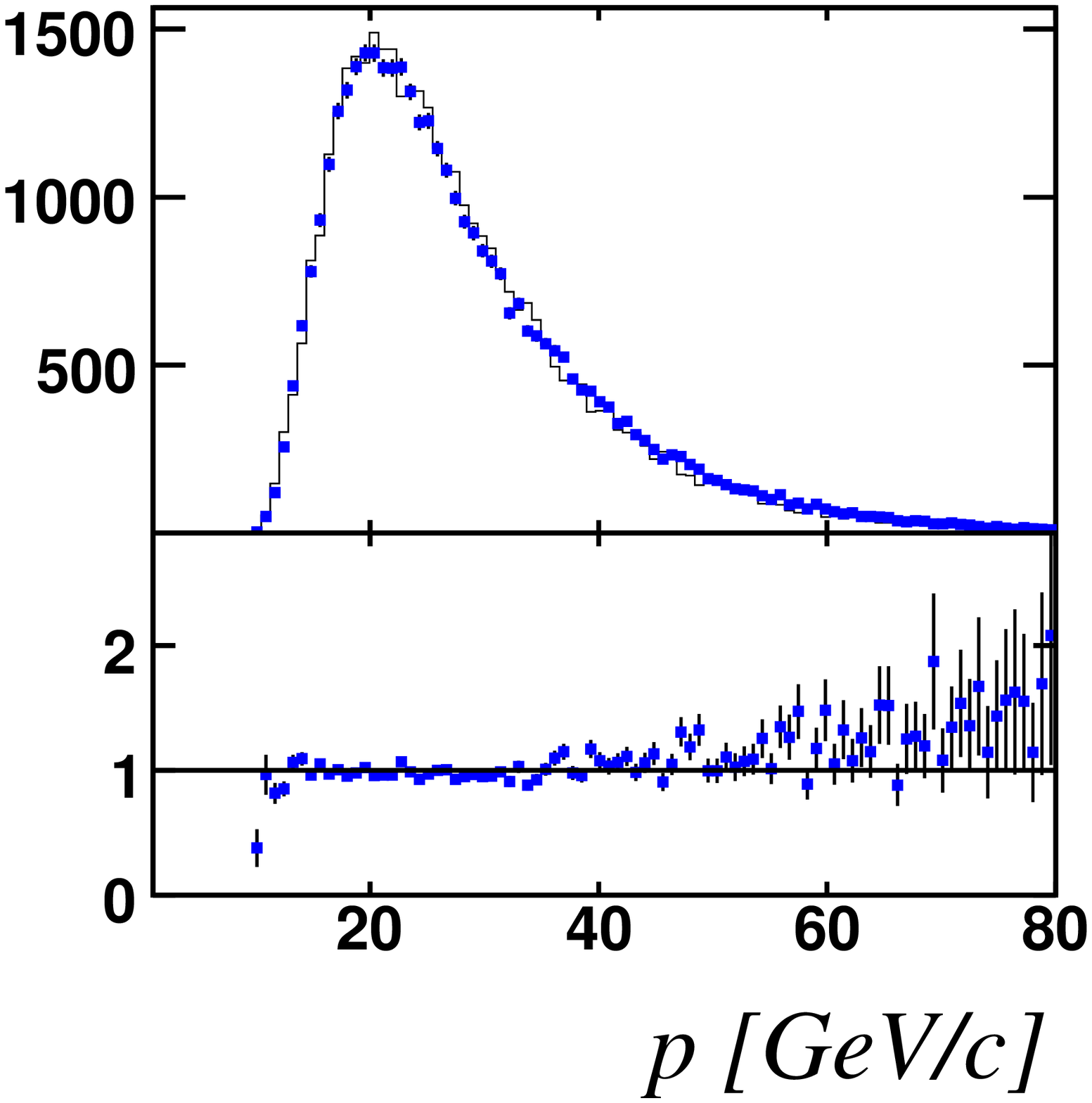}
\includegraphics[width=0.25\textwidth]{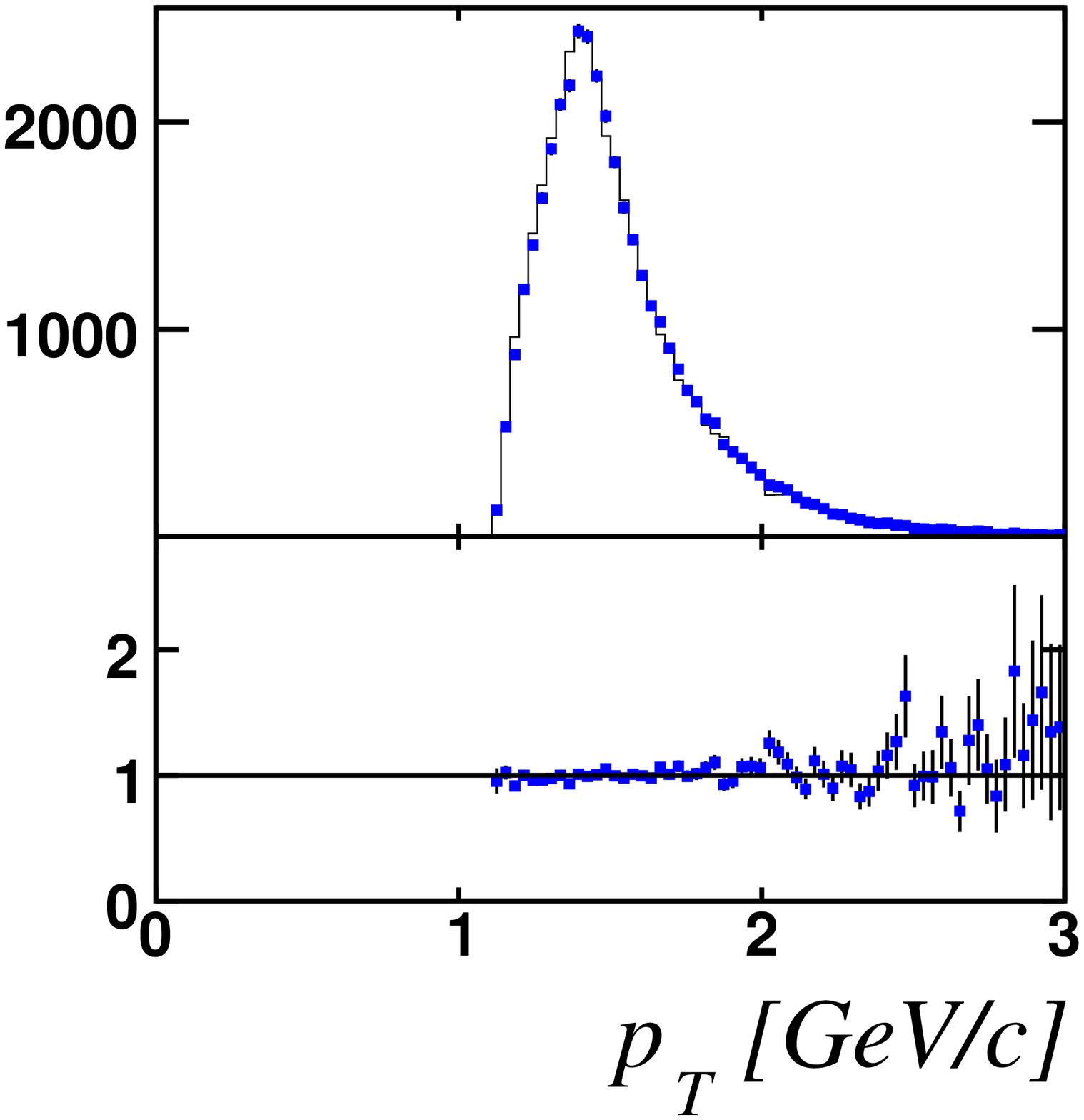}}
\caption{Comparison between data and MC simulations -- The
  distributions and ratios of Data/MC for: kinematic variables:
  $Q^{2},\, y$ (1st row). Total and transverse momentum $p$ and
  $p_{T}$ of the high $p_T$ hadron (2nd raw).}
\label{fig:lowq2mcrealcomp}
\end{wrapfigure}

The reason for splitting the $Q^2$ range in two
complementary regions is that for the low $Q^2$ region the resolved
photon contributions are considerably higher ($\approx 50 \ \%$) than in
the high $Q^2$ region, which contains practically only the three processes previously mentioned. This means that the QCD hard scale is also different for both $Q^2$ regions: for low $Q^2$ the scale is given by the high $p_T$ hadrons, while for the high $Q^2$ is given by the $Q^2$ value itself.  

A more complicated description of the physics than pure QCD
in lowest order needs to be included in the MC simulation for this
case. Therefore the event generator used in this analysis is PYTHIA
6.2 \cite{Sjostrand:2000wi} which covers the physiscal processes for quasi-real photoproduction.

In this analysis the selection is essentially the
same as in high $Q^2$ region plus a slightly strict set of cuts:
$x_F>0.1$, $z > 0.1$, and $\sum p_T^2 >2.5 \ (\mbox{GeV}/c)^2$. The
data sample in this region is 90 \% of the whole data for all
$Q^2$ range. The weighting
method used in the high $Q^2$ analysis is not applied in this
case. 

The MC simulated and real data samples of high $p_{T}$ events are
compared in Fig.~\ref{fig:lowq2mcrealcomp} for $Q^2$, $y$ (1st row), and for the total
and transverse momenta of the high $p_T$ hadron (2nd row), showing a good
agreement.

The gluon polarisation in the low $Q^2$ region is extracted using averaged values as shown by this expression:

\begin{eqnarray}
  \nonumber \left \langle \frac{A_{\rm LL}}{D} \right \rangle &=&
  R_{\rm PGF} 
            \left\langle \frac{\hat a_{\rm LL}^{\rm PGF}}{D} \right\rangle
\left \langle \frac{\Delta G}{G} \right \rangle  
+ R_{\rm QCDC} \left \langle \frac{\hat a_{\rm LL}^{\rm QCDC}}{D}
  A_1 \right \rangle  
+  \sum_{f,f^\gamma}
R_{ff^\gamma} \left\langle \hat a_{\rm LL}^{ff^\gamma}  \frac{\Delta
f}{f} \frac{\Delta f^\gamma}{f^\gamma} \right\rangle \\
  &&
\label{eq:allcontribs}
\end{eqnarray}

$R_{ff^\gamma}$ is the fraction of events in the high $p_{T}$
sample for which a parton $f$ from the nucleon interacts with a parton
$f^\gamma$ from a resolved photon; $A_1$ is the virtual photon deuteron asymmetry measured in an inclusive sample; $\Delta f/f$ ($\Delta f^\gamma/f^\gamma$) is the polarisation of quarks
or gluons in the deuteron (photon).

This analysis was performed using a data
sample from the years 2002 to 2004. For more details about this analysis the reader is invited
to look into ref.~\cite{Ageev:2005pq}.

\section{Results}
\label{res}
The preliminary measurements of the gluon polarisation in low and high $Q^2$
regions, using data from the years 2002 to 2004, are:

\begin{eqnarray*}
  \left(\Delta G/G \right)_{{\rm low} \ Q^2} &=& 0.02 \pm 0.06_{(stat.)} \pm
  0.06_{(syst.)} \qquad {\rm with} \quad x_G= 0.09^{+0.07}_{-0.04}\\
  \left(\Delta G/G \right)_{{\rm high} \ Q^2} &=& 0.08 \pm 0.10_{(stat.)} \pm 0.05_{(syst.)} \qquad {\rm with} \quad x_G= 0.08^{+0.04}_{-0.03} 
\end{eqnarray*}

The average of the hard scale, $\mu^2$, for low and high $Q^2$ is
about 3
$(\mbox{GeV}/c)^2$. $x_G$ is the momentum fraction carried by
the probed gluons obtained from the MC parton kinematics. The result of the measurement for low $Q^2$ using data from
2002 and 2003 can be found in~\cite{Ageev:2005pq}.

\begin{wrapfigure}[21]{r}{8cm}
\vspace{-10pt}
\centerline{
\includegraphics[width=0.5\textwidth]{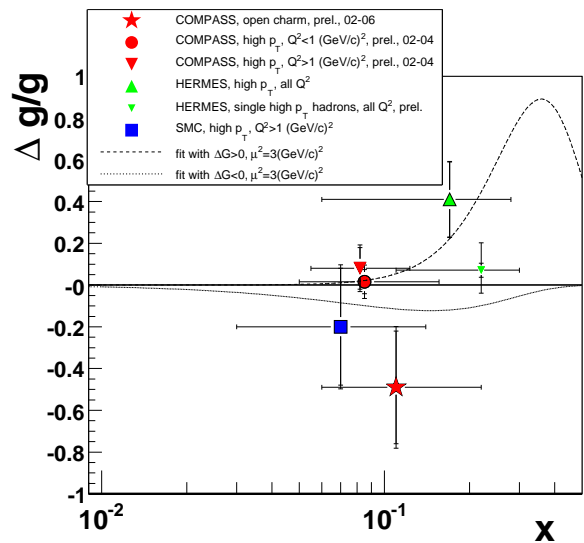}}
\caption{Comparison of $\Delta G/G$ measurements from COMPASS \cite{comp.del_sigma}, SMC
  \cite{Adeva:2004dh}, and  HERMES \cite{Airapetian:1999ib}. The two
  curves correspond to the parametrisation from the NLO QCD analysis
  in the $\overline{MS}$ scheme with scale at 3 $(\mbox{GeV/}c)^2$.}
\label{fig:dgg}
\end{wrapfigure}

Fig.~\ref{fig:dgg} shows these new values of $\Delta G/G$ together with the
preliminary value from the open charm analysis. Also the figure shows the measurements
from SMC collaboration, from the high $p_T$ analysis for
the $Q^2 >1 \ (\mbox{GeV/}c)^2$ region \cite{Adeva:2004dh} and also the measurements from
HERMES collaboration, for single hadrons and high $p_T$ hadron pairs
analyses \cite{Airapetian:1999ib}. The curves in the figure are the parametrisation of $\Delta
G/G (x)$ using a NLO QCD analysis done by COMPASS
\cite{comp.del_sigma} in the $\overline{MS}$ scheme with a
renormalisation scale $\langle \mu^2 \rangle = 3 \ (\mbox{GeV/}c)^2$. The dashed line curve is the QCD fit assuming that $\Delta G > 0$, the
dotted line is the  QCD fit assuming $\Delta G < 0$. It is seen that
both results from high $p_T$ analyses, for high and low $Q^2$ regions, are
compatible with each other and also, within their $x_G$ region, in
agreement with the NLO QCD fits.

\section{Conclusions}

The preliminary values of the gluon polarisation for low and high
$Q^2$ regions were presented. The gluons were probed at an average
scale $\langle \mu^2 \rangle \approx 3 \ (\mbox{GeV/}c)^2$. Both measurements show that
the gluon contribution to the nucleon spin for $x_G \approx 0.1$ is compatible
with zero. In that region of $x_G$ the presented measurements are in
agreement with other well known results.

\end{document}